\documentclass{aa}  
\usepackage{dsfont}
\usepackage{graphicx}
\usepackage{txfonts}
\usepackage{lipsum}
\usepackage{subcaption}
\usepackage{lscape}
\usepackage{placeins}

\usepackage{rotating}
\usepackage{graphicx}
\usepackage{sidecap}
\usepackage{subcaption}
\usepackage{url}
\usepackage[colorlinks=true]{hyperref}
\usepackage{soul}
\hypersetup{allcolors=blue}
\usepackage[normalem]{ulem}
\usepackage[kerning=true, tracking=true]{microtype}
\usepackage{lipsum}
\usepackage{txfonts}
\usepackage[]{xcolor}
\usepackage{float}

\bibpunct{(}{)}{;}{a}{}{,}
\DeclareUnicodeCharacter{00A0}{~}

\newcommand{\bifrost}{{\texttt{Bifrost}}}

\begin{document}

\title{Accelerating 3D Non-LTE Synthesis with Graph Neural Networks}

\titlerunning{Fast 3D Non-LTE Populations with GNNs}
\authorrunning{Vicente Ar\'evalo et al.}

\author{A. Vicente Ar\'evalo \inst{1}, A. Asensio Ramos\inst{2}, C. J. D\'iaz Baso\inst{3,4}}

\institute{Institut für Sonnenphysik (KIS), Georges-K\"ohler-Allee 401A, 79110 Freiburg, Germany
\and Instituto de Astrofísica de Canarias, C/Vía Láctea s/n, E-38205 La Laguna, Tenerife, Spain
\and Institute of Theoretical Astrophysics,
University of Oslo, %
P.O. Box 1029 Blindern, N-0315 Oslo, Norway
\and
Rosseland Centre for Solar Physics,
University of Oslo, %
P.O. Box 1029 Blindern, N-0315 Oslo, Norway
}
 
   \abstract{Spectropolarimetric interpretation of chromospheric lines requires solving the radiative transfer problem under non-local thermodynamic equilibrium (non-LTE) conditions.
    The atomic level populations must be computed self-consistently with the radiation field.
    Traditional inversion codes employ 1.5D approximations for computational tractability, neglecting horizontal radiative transfer that becomes significant near magnetic structures and dynamic chromospheric features.}
    {We present a method to solve the 3D statistical equilibrium (SE) populations of any atom using Graph Neural Networks (GNNs), extending prior work in 1.5D to the full 3D domain.
    The goal is to develop a fast surrogate model that fully accounts for horizontal and vertical radiative coupling while being robust to changes in the underlying grid, enabling future 3D non-LTE inversions.}
    {We discretize the solar atmosphere as a directed graph, where nodes encode local physical properties (temperature, velocity, magnetic field, electron density, etc.) and edges encode geometric distances between spatial points.
    An Encode-Process-Decode GNN architecture propagates information across the 3D domain to capture the radiative coupling efficiently.
    The network is trained on a \bifrost\ radiation-MHD simulation using \ion{Ca}{ii} populations computed with \texttt{Multi3D} as ground truth.}
    {The trained GNN accurately predicts populations of the five-level \ion{Ca}{ii} atom plus continuum.
    Correlations exceed $0.99$ in the photosphere and mid-chromosphere; errors increase in the upper chromosphere, where non-local effects dominate but remain unbiased with Gaussian distributions.
    Inference is  $\sim10^6$ faster than traditional iterative SE solvers.
    Spectral synthesis of the \ion{Ca}{ii}~8542\,\AA\ line yields intensity profiles with $\lesssim 2\%$ mean residuals relative to the full 3D solution, opening
    up the option of more realistic inversions of strong chromospheric lines.}
    {The 3D GNN framework bypasses the computational bottleneck of iterative SE solvers while preserving essential non-LTE physics, including horizontal radiative transfer.
    The method is readily extensible to other atoms and paves the way toward routine 3D non-LTE inversions.}


\maketitle
\nolinenumbers
\section{Introduction}
\label{sec:introduction}

The accurate modeling of solar chromospheric spectra requires a rigorous treatment of non-Local Thermodynamic Equilibrium (non-LTE) radiative transfer.
In these tenuous layers, the plasma density is too low for collisions to enforce thermal equilibrium, and the atomic level populations become strongly coupled to the radiation field.
Since photons can travel large distances before being absorbed, the state of the plasma at any given point depends on the physical conditions across a vast surrounding volume.
This non-local coupling necessitates the simultaneous solution of the Radiative Transfer (RT) equation and the Statistical Equilibrium (SE) equations in full 3D geometry.

Several massively parallel codes have been developed to tackle this computationally intensive problem, including \texttt{Multi3D} \citep{Leenaarts2009}, \texttt{PHOENIX/3D} \citep{2010A&A...509A..36H}, and \texttt{PORTA} \citep{2013A&A...557A.143S}.
These codes typically employ iterative schemes, such as Multi-level Accelerated Lambda Iteration \citep[MALI;][]{Rybicki_Hummer_1991}, to find a self-consistent solution for the radiation field and atomic populations.
While they provide the gold standard for synthetic spectra, their high computational cost (often requiring up to millions of CPU hours per snapshot) limits their application to a small number of static snapshots or simplified test cases.

This computational bottleneck is one of the primary obstacles for 3D spectropolarimetric inversions.
Inversion codes like \texttt{NICOLE} \citep{SocasNavarro2015}, \texttt{SNAPI} \citep{Milic2018}, \texttt{DeSIRe} \citep{desire_inpress} or \texttt{STiC} \citep{2019A&A...623A..74D} infer atmospheric properties by iteratively fitting synthetic spectra to observations.
Because they must execute the forward model hundreds of times per pixel, they are forced to rely on the 1.5D approximation, treating each pixel as an independent column and neglecting horizontal radiative transfer.
However, this assumption breaks down in the dynamic chromosphere, where structures like fibrils and shock fronts exhibit strong horizontal gradients.
Ignoring 3D radiative effects can introduce significant biases in the inferred temperature and magnetic field stratification \citep{polaris_original, Jaume_Bestard_2021, biorgen}.
Furthermore, 3D radiative transfer introduces non-local scattering effects that smooth the source function spatially and are essential for reproducing characteristic chromospheric features such as H$\alpha$ fibrils, which are notably absent in 1.5D synthesis \citep{Leenaarts2012}.

To make 3D inversions tractable, one would need a forward model orders of magnitude faster than current iterative solvers. Ideally, this forward model should also be differentiable to seamlessly incorporate it into standard optimization methods used in inversions \citep[e.g.,][]{2025A&A...693A.170D}.

Recent advances in deep learning suggest a way forward. \citet{Chappell_Pereira_2022} pioneered this approach with \emph{SunnyNet}, a Convolutional Neural Network (CNN) that predicts 3D non-LTE populations for H$\alpha$ with a speedup of $\sim 10^5$ over traditional codes. While groundbreaking, CNNs rely on regular Euclidean grids where spatial information is implicitly encoded in the pixel spacing. This can limit their generalization to observations or simulations with different spatial resolutions or non-uniform grids. We are aiming to remove that grid dependence while allowing for arbitrary graph topologies. To enable routine 3D inversions across multiple instruments with varying pixel scales, the forward model must be able to handle arbitrary spatial sampling without requiring retraining.

Graph Neural Networks (GNNs) offer a more geometrically robust alternative.
As demonstrated by \citet{Andreu_2022} in the 1.5D regime, GNNs treat the atmosphere as a flexible graph where nodes represent spatial points, and edges encode the physical distance explicitly.
This decouples the learning process from the grid topology: the network learns the physics of radiative propagation with distance rather than patterns on a fixed pixel array.

In this work, we extend this GNN framework to the full 3D domain for the \ion{Ca}{ii} atom as our primary target. The \ion{Ca}{ii} infrared triplet (8498, 8542, 8662\,\AA) serves as an ideal testbed; these lines are commonly used chromospheric diagnostics in modern solar telescopes due to their strong sensitivity to local temperature and magnetic fields \citep[e.g.,][]{2017A&A...599A.133A,2019A&A...623A.178D,2021A&A...649A.106Y}. Crucially, their formation is influenced by 3D radiative transfer effects. 
By explicitly modeling the 3D spatial connectivity, our architecture provides a fast, geometry-agnostic surrogate for the SE equations that naturally accounts for horizontal radiative transfer, paving the way for routine 3D non-LTE inversions.

\section{Theoretical background}
\label{sec:theory}

In the solar chromosphere, the low plasma density decouples the radiation field from local thermal conditions, invalidating the assumption of Local Thermodynamic Equilibrium (LTE).
To accurately model this regime, one must determine the atomic populations $\vec{n}$ by solving the equations of Statistical Equilibrium (SE):

\begin{equation}
\label{eq:SEE}
\sum_{j \neq i} n_j P_{ji} - n_i \sum_{j \neq i} P_{ij} = 0,
\end{equation}

\noindent where the first term represents transitions into level $i$ from all other levels $j$, and the second term represents transitions out of level $i$.
The total transition rate $P_{ij} = C_{ij} + R_{ij}$ includes both local collisional terms ($C_{ij}$) and non-local radiative rates ($R_{ij}$).
Each radiative rate depends on the mean intensity $\bar{J}$, which is the integral of the specific intensity $I_\lambda$ over all angles and frequencies of that transition.
Specifically, $R_{ij} = B_{ij}\bar{J}_{\nu_{ij}}$ for radiative excitation (where $B_{ij}$ is the Einstein coefficient), making the rates inherently non-local since $\bar{J}$ depends on the radiation field from distant atmospheric layers.
For resonance lines like \ion{Ca}{ii} 8542~$\AA$, scattering typically dominates over pure absorption in the chromosphere. This means photons can be scattered multiple times across large spatial scales before being thermalized, creating strong coupling between spatially separated atmospheric regions—the essence of the 3D non-LTE problem.

At the same time, the specific intensity is governed by the radiative transfer equation:

\begin{equation}
\label{eq:rte}
\frac{\mathrm{d} I_\lambda}{\mathrm{d} \tau_\lambda} = I_\lambda - S_\lambda,
\end{equation}

\noindent where the source function $S_\lambda \equiv \varepsilon_\lambda / \eta_\lambda$ is the ratio of emissivity to absorption, and $\tau_\lambda$ is the optical depth of the computed ray.
Because $\varepsilon_\lambda$ and $\eta_\lambda$ depend linearly on the populations $\vec{n}$, Eqs.~(\ref{eq:SEE}) and (\ref{eq:rte}) form a coupled, non-linear system: the level populations determine the radiation field, which, in turn, determines the populations.

Traditional 3D non-LTE codes (e.g., \texttt{Multi3D}) solve this circular dependency via computationally expensive iterative schemes like the accelerated $\Lambda$-iteration.
However, we note that if we had a method to instantaneously know non-LTE populations, the emergent spectrum can be synthesized in a single step by applying the formal solution of Eq.~(\ref{eq:rte}).
Additionally, such a single formal solution can be trivially parallelized, contrary to the full non-LTE problem.
As a consequence, having a fast predictor of the level populations for the given physical conditions can potentially accelerate non-LTE synthesis by a large margin.
We propose to bypass the iterative bottleneck by using a GNN to 
produce level populations at every position in the 3D model, leveraging information in the surroundings of the point.
This allows us to predict self-consistent populations directly and compute spectra via a single formal solution.

The following section describes the specific GNN architecture and graph construction strategy employed to efficiently approximate this population predictor while maintaining computational tractability for 3D domains.

\section{Methodology and architecture}
\label{sec:Architecture}

Building on the success of \citet{Andreu_2022}, who demonstrated that GNNs can accurately predict 1.5D non-LTE populations with orders-of-magnitude speedup while maintaining grid flexibility, we extend this approach to map the physical parameters of the three-dimensional solar atmosphere to the level populations of a \ion{Ca}{ii} model atom.
While the previous approach modeled 1.5D semi-infinite atmospheres using optical depth as a distance metric, the 3D nature of the current problem necessitates 
significant changes in graph construction, dimensionality, connectivity, and encoded parameters to capture horizontal radiative transfer effects while maintaining 
computational efficiency. We describe them in the following.

\subsection{The GNN framework}
\label{subsec:graphnet_framework}
Using the same approach followed by \citet{Andreu_2022}, we discretize the solar atmosphere not as a regular Cartesian grid, but as a directed graph where edges encode information flow from surrounding context nodes toward the central column being inferred.
The set of nodes of the graph represents the properties of the atmosphere (either physical properties or the inferred populations) at 
particular points, and the set of edges represents the radiative coupling between them.
This graph-based representation gives flexibility to handle arbitrary geometries and essential non-local dependencies for the non-LTE problem.

We adopt the Encode-Process-Decode architecture \citep{bataglia2018}, implemented using the \textsc{Pytorch} library \citep{PyTorch}. 
The inference pipeline proceeds in three distinct stages (for a visual 1D representation we refer the reader to Fig. 1 of \cite{Andreu_2022}):

\noindent\textbf{Encoder}: The physical properties at each node and the geometric relationships between them (edges) are projected into a higher-dimensional latent space.
Two independent Multi-Layer Perceptrons (MLPs), $f_{v}$ and $f_{e}$, transform the input feature vectors $\vec{p}_i$ and edge attributes $\vec{x}_{ij}$ into latent representations $\vec{v}_i$ and $\vec{e}_{ij}$ of dimension $d=128$:
$$
\vec{v}_i = f_{v}( \vec{p}_i ) \quad \vec{e}_{ij} = f_{e}( \vec{x}_{ij} )
$$
    
\noindent\textbf{Process}: The core computation occurs through $N$ message-passing steps, allowing information to propagate across the domain.
In this work, we use $N=16$ steps.
At each step $t$, edge and node states are updated sequentially.
First, an edge model $f_{E}^{t}$ updates the edge features based on the concatenated states of the
connected nodes and the edge, with a residual connection:
$$
\vec{e}_{ij}^{t+1} = \vec{e}_{ij}^{t} + f_{E}^{t+1}(\vec{e}_{ij}, \vec{v}_i, \vec{v}_j)
$$
This edge model is another MLP that is shared for all the edges in the graph, but different for each message-passing step.
Subsequently, an aggregator model $f_A$ aggregates the incoming "messages" (updated edge information) from all neighboring nodes $v_k^t$ into one message $\vec{e}_{j}^{t+1}$ to then use it to update the node with another MLP model $f_{V}$:
$$
\vec{e}_{j}^{t+1} = \sum_k\ f_{A}^{t+1}(\vec{v}_{k}^t, \vec{e}_{kj}^{t+1}) \rightarrow \vec{v}_{j}^{t+1} = \vec{v}_j^t + f_V^{t+1} (\vec{v}_j^t, \vec{e}_j^{t+1}),
$$
where $\sum$ represents a permutation-invariant aggregation function (in our case \texttt{scatter\_mean} to reduce gradient explosion\footnote{By gradient explosion we are referring to the numerical instability occurring when large error gradients accumulate and cause divergent weight updates when back-propagating.} when back-propagating \footnote{The recursive application of the chain rule to compute objective function gradients to optimize the weights of the model.} during training).
    
Like $f_E^t$, the $f_A^t$ and $f_V^t$ models are shared across all nodes in the graph, but differ at each message-passing step.
This mechanism mimics the physical propagation of radiation: the message-passing steps literally propagate the radiation field through the atmospheric graph, where the local state of a node is iteratively updated by the incoming radiation (messages) from neighboring regions.
The choice of $N=16$ message-passing steps represents a significant reduction compared to the original 1.5D work, where $N\sim100$ was required \citep{Andreu_2022}.
We attribute this to the richer 3D connectivity, which allows each step to capture more context and physically meaningful interactions, thereby reducing the required network depth and total parameter count.\\

\noindent\textbf{Decoder}: Finally, a decoder MLP $f_n$ projects the final latent node states back into the physical space.
The output is the normalized population of each atomic level of \ion{Ca}{ii} (five bound levels plus \ion{Ca}{iii}):
$$
\vec{n}_i = f_n(\vec{v}_i^N)
$$

\subsection{Graph Construction}
\label{subsec:graph_construction}

Each node $i$ is initialized with an 11-dimensional feature vector $\vec{p}_i$ containing the local thermodynamic and magnetic properties: temperature ($T$), 
velocity vector in Cartesian coordinates ($\vec{v}$), magnetic field vector also in Cartesian coordinates ($\vec{B}$), number densities of electrons ($n_{\rm e}$), gas pressure ($p_{gas}$), and mass density ($\rho$).
Crucially, we also include the vertical geometric height $z$ as an explicit input feature to help the network learn height-dependent stratifications.

A major departure from the 1.5D approach is the definition of edge attributes. In 1.5D, the optical depth distance $\Delta\tau$ naturally encodes opacity.
However, calculating 3D optical distances requires solving the same non-LTE problem we aim to bypass.
For that reason, we encode the geometric Euclidean distance $|\vec{r}_{ij}|$ in the edges.
While less physically informative than optical depths $\Delta\tau$, the network learns to infer opacity structures effectively from the combination of geometric distance and local densities provided in the nodes.

Transitioning from 1.5D to 3D requires a fundamental redefinition of the graph topology.
In 1.5D, nodes were sequentially connected along the vertical optical depth scale, with only one connection to and from previous and next nodes.
In 3D, we must account for lateral radiation transfer and adopt a more elaborate graph-building strategy.

\begin{figure}[t]
    \centering
    \includegraphics[width=1.\linewidth]{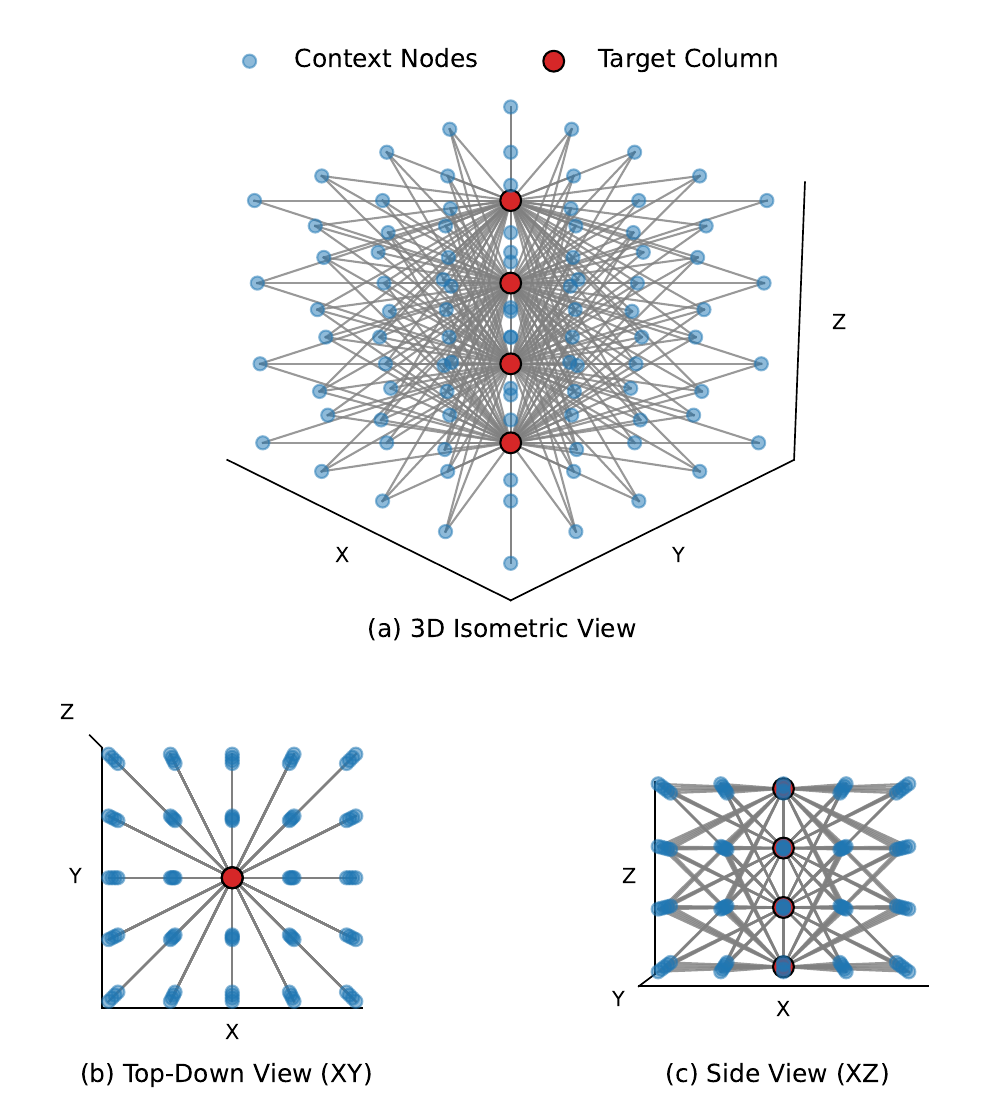}
    \caption{Schematic representation of the 3D graph topology and connectivity scheme.
    Red nodes indicate the central column where atomic populations are inferred, while blue nodes represent the surrounding physical context required to capture non-local 3D radiative transfer effects.
    This illustration corresponds to $\pm 2$ neighbors, stride $S=1$, and radii $R=3$.
    (a) Isometric 3D view of a sample subvolume.
    (b) Top-down (XY) projection illustrating the ``star-like'' connectivity.
    (c) Side (XZ) projection showing the vertical stratification and how radii $R$ and distance to the center affects the connectivity.
    }
    \label{fig:connectivity}
\end{figure}

A fully connected graph within the subvolume would grow as $O(n^2)$, where $n$ is the number of nodes considered in the graph, 
leading to excessive memory consumption even for small neighborhoods.
To address this, we adopt a strategy focused on inferring populations only for the central column of each subvolume, which benefits from the richest context.
We define the stride $S$ (the sampling interval along x and y) and connectivity radius $R$ (maximum distance for edge connections) before describing the pruned connectivity scheme below.
Since our primary goal is to infer the populations for the central column, we implement a pruned connectivity scheme, retaining only those edges that originate or terminate at a node within the central column.
Figure~\ref{fig:connectivity} shows a sample graph to illustrate the graph topology.
This ``star-like'' connectivity significantly reduces the graph density, while preserving the direct radiative coupling between the target node and its 3D environment.
This optimization allows us to process larger context volumes during training and inference without exceeding GPU memory limits.

\section{Training}
\label{sec:training}

\subsection{Training data}
For training the graph neural network, a dataset with fully known physical parameters and a 3D radiative transfer solution is required.
We use a snapshot of the ``enhanced network'' simulation carried out with the \bifrost\ radiation-MHD code \citep{Carlsson_2015}
.
The snapshot provides 3D volumes of the physical quantities, together with the non-LTE populations computed using \texttt{Multi3D} \citep{Leenaarts2009} for a five-level \ion{Ca}{ii} model atom plus continuum treated in complete frequency redistribution.
The simulation domain spans $24 \times 24 \times 14.4~\mathrm{Mm}^3$ (from the photosphere up to the corona), discretized on 
a grid of $504 \times 504 \times 496$ points.
As already described, we used $T$, $\vec{v}$, $\vec{B}$, $n_{\rm e}$, $p_{gas}$, $\rho$, and $z$ as input nodes in the GNN, 
while the \ion{Ca}{ii} level populations $n_i$ ($i=0,\ldots,5$) are used as outputs.
The six levels correspond to the five bound states of \ion{Ca}{ii} ($4^2S_{1/2}$, $3^2D_{3/2}$, $3^2D_{5/2}$, $4^2P_{1/2}$, $4^2P_{3/2}$) plus the \ion{Ca}{iii} ground state.
Feature scaling is critical for the proper convergence of the neural model since 
some of the input variables and all output variables span more than 10 orders of magnitude from the photosphere to the high chromosphere. 
We apply distinct normalization strategies for the input physical features $\vec{p}_i$ and the output populations $\vec{n}_i$:\\

\noindent\textbf{Input features}: Variables with relatively linear behavior (e.g., temperature) are standardized to zero mean and unit variance. Quantities that span several orders of magnitude, such as densities, are first transformed to a logarithmic space (after dividing by the mean $\mu_x$) before rescaling: $\hat{x} = \log_{10}(x/\mu_x)$. Magnetic field and velocity components are scaled with the normal scaler as $\hat{x} = x/\sigma_x$, with $\sigma_x$ being the standard deviation of the feature. Height ($z$) is given in Mm without normalization.\\

\noindent\textbf{Target populations}: The atomic populations are normalized using a logarithmic transformation. We first compute the fractional population relative to the total calcium abundance, then compress the dynamic range:
\begin{equation}
    n^{\rm frac}_i = \frac{n_i}{\sum_j n_j}
    \quad\rightarrow\quad
    \hat{n}^{\rm frac}_i = \frac{1}{k}\log_{10}\left(\frac{n^{\rm frac}_i}{\mu(n^{\rm frac}_i)}\right)
\end{equation}
where $k=4.0$ and $\mu$ is the mean over the feature\footnote{The $k$ value is chosen empirically to approximately normalize the output variance to unity}. This transformation emphasizes variations in minority-level populations (common in the upper part of the atmosphere) without losing information about the dominant levels.\\

A point to note here is that as physical quantities span several orders of magnitude from the photosphere to the corona, normalizing with a global $\mu$ and $\sigma$ would bias the normalization toward lower, denser regions.
To capture the stratification properly, the normalizations are calculated with the mean and variance at each specific height ($\mu(z)$, $\sigma(z)$). This happens in the data preparation stage (before training) and is done with the statistics of the full dataset.

Training on the entire simulation domain simultaneously is not optimal (i.e., unmanageable graph size, local minima avoidance from mini-batch training, speed of the minimization, \ldots).
Instead, we construct graphs from subvolumes centered on the target column.
We select neighbors within a defined range (e.g., $\pm 4$ columns in x and y) from the central column, and to improve the network's ability 
to handle different spatial scales and reduce the graph size, we employ a striding technique during training.
That means we randomly select a stride $S \in \{1,2,3,4,5\}$ and subsample the grid, skipping $S$ nodes.
The connectivity radius $R$ is dynamically adapted based on the stride to maintain physical context $R = R_0\frac{2S}{3}$, with $R_0 = 5$.

The training graphs were therefore constructed using the following steps:
\begin{itemize}
    \item We select one central column where the populations are inferred.
    \item We randomly chose a stride size $S$ which will determine the spatial subsampling of the original grid.
    \item We select $N$ neighboring columns in each direction, where the separation between adjacent columns will be determined by the stride, effectively having a $((2N+1) \times (2N+1) \times N_z)$ nodes in the graph. The $N$ is fixed in our architecture and is chosen to be $N=4$. So our graphs always have $11 \times 11 \times 55$ nodes, regardless of the physical volume spanned.
    \item We construct a graph, connecting the nodes within the dynamically computed radii $R$.
    \item Finally, we remove the connections that do not start or end at a node in the central column as described in Sec. \ref{subsec:graph_construction}
\end{itemize}

These graphs obtained for subvolumes are separated into training and testing. The testing is obtained from the region $x=[358,484]$ and $y=[358,484]$. This corresponds
to $\sim 8\%$ of the available columns, which are never used during training.
It is worth noticing that, given the fact that we are only predicting the central column of the graph, and due to the striding technique, we are losing 
some columns close to the edges. Since the graphs with $S=4$ and $N=4$ use columns 16 pixels away from the central column, we therefore remove from the training all 
columns with $x, y \le 16$ and $x, y \ge 488$. 

\subsection{Loss function and optimization}
\label{subsec:loss_normalization}
We use a standard mean squared error (MSE) loss function on the transformed values.
Training was performed on a single NVIDIA H100 GPU and required approximately 50 hours.
We used a batch size of 16 subgraphs and trained for 120 epochs.
The training and validation loss curves exhibit the expected behavior: the training loss decreases monotonically while the validation loss gets flat after approximately 40 epochs.
The best model was selected based on the lowest validation loss to prevent overfitting.

We use the Adam optimizer \citep{adam14} with an initial learning rate of $3\times10^{-4}$.
To ensure convergence to a precise solution, we employ a multi-step learning rate scheduler.
The learning rate is decayed by a factor of $\gamma = 0.5$ at specific milestones (epochs 20, 40, 90, and 115), allowing the network to settle into a minimum after an initial exploration phase.
We point out that additional physics-informed terms could be included in the loss function to better constrain the solution.
For example, one could add a regularization to enforce $\sum_i n_i = n_{T}$. We preferred to 
train simply with the MSE and leave the effect of physics-informed terms for a later study.

\section{Inference}
\label{sec:Inference}

\begin{figure*}[htbp]
    \centering
    \includegraphics[width=0.68\linewidth]{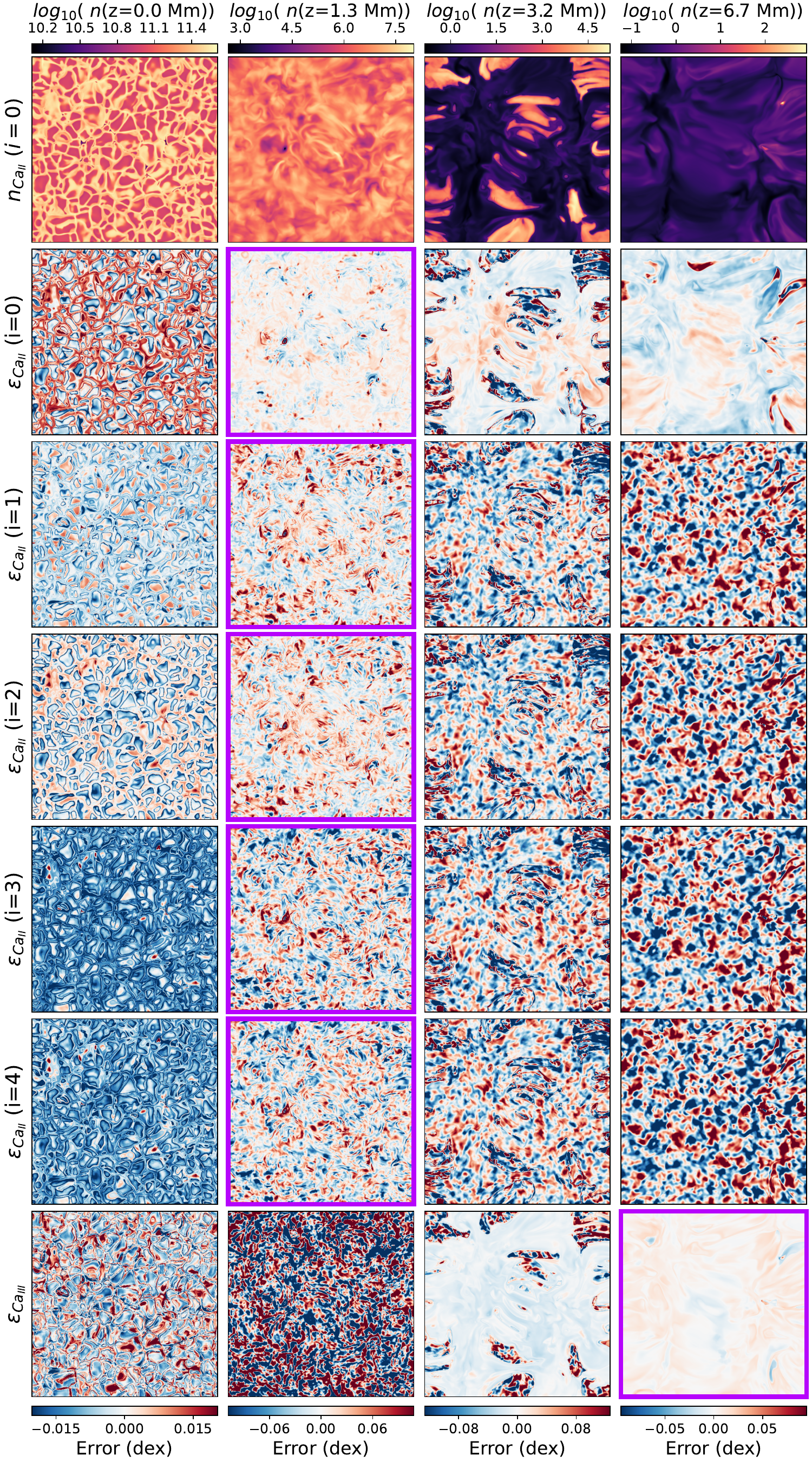}
    \caption{Spatial maps of the inference errors of the predicted \ion{Ca}{ii} populations at different atmospheric heights $z \sim [0, 1, 3, 7 ]$ Mm.
    Top row: ground-level \ion{Ca}{ii} populations at various heights for context. Second-to-Bottom rows: error maps between the GNN predictions and the \texttt{Multi3D} ground truth.
    The color scale shows the errors on the dex scale ($log(x_{pred}) - log(x_{true})$).
    Columns correspond to increasing height from the photosphere (left) to the upper chromosphere (right).
    Magenta borders highlight the height at which each atomic level reaches its peak population, indicating where that level dominates the ionization balance.}
    \label{fig:inference_maps}
\end{figure*}

\begin{figure}[t]
    \centering
    \includegraphics[width=\columnwidth]{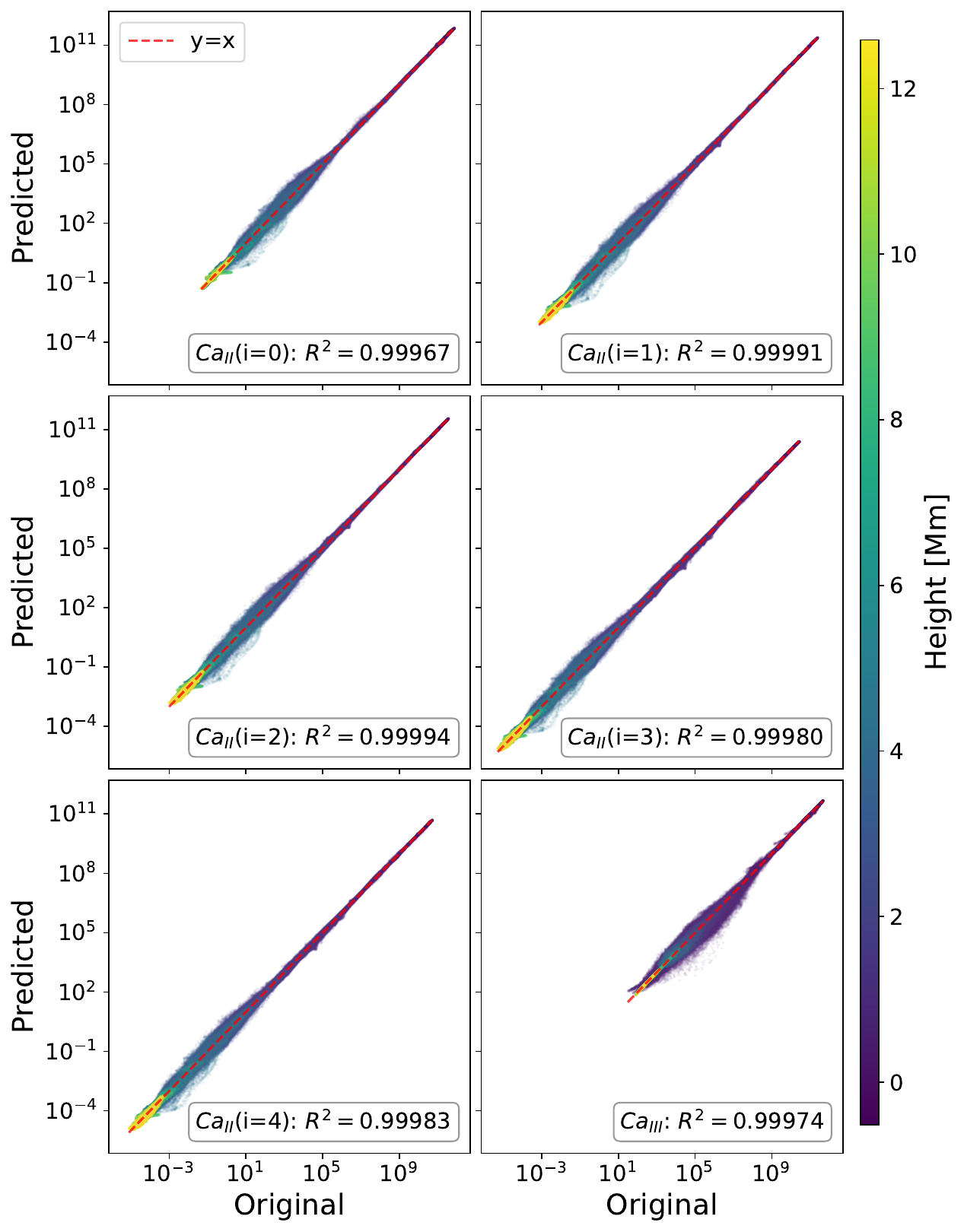}
    \caption{Scatter plots of predicted versus ground-truth \ion{Ca}{ii} level populations, color-coded by geometric height. High density along the diagonal indicates excellent correlation.}
    \label{fig:inference_scatter}
\end{figure}

Once the training is complete, we evaluate the model's inference capabilities with a fixed stride of $S=4$.
We obtain high-fidelity reconstruction on the atomic population maps throughout the entire volume.
For completeness, the error estimation of our model has also been done in this entire volume reconstruction, but identical statistical properties are found when tested only in the testing set.
As illustrated in Fig.~\ref{fig:inference_maps}, the inferred populations capture the complex spatial morphology throughout the full solar atmosphere, reproducing the granulation patterns in the lower photosphere and the intricate fibril structures in the chromosphere. 
A quantitative assessment of the correlation plots (Fig.~\ref{fig:inference_scatter}) shows that the model achieves correlations of $>0.99$ for all the levels throughout the photosphere and mid-chromosphere. Furthermore, the error maps (Fig.~\ref{fig:inference_maps})
reveal that the majority of the inference error is concentrated in the higher layers of the computational domain for the \ion{Ca}{ii} populations, and toward the lower layers for the \ion{Ca}{iii}.
As a visual aid, we marked with magenta the edges of the panels at the heights at which the relative population is largest for each atomic level.

We attribute the observed behavior to two primary factors:
\begin{itemize}
    \item Relative error sensitivity: The absolute population densities in the upper atmosphere drop by several orders of magnitude. Consequently, even small absolute deviations in the network's prediction result in larger relative errors when compared to the dense, lower layers.
    The inverse can be said about \ion{Ca}{iii}, which is relatively more abundant in the upper region and magnifies the errors in the lower layers.
    \item Physics complexity: The upper layers are optically thinner, and non-LTE and 3D RT effects are more pronounced. Therefore, the mapping between physical conditions and atomic populations is significantly more non-local and non-linear at these heights, presenting a harder challenge for the network compared to deeper regions, where
    populations are closer to LTE.
\end{itemize}

\subsection{Error estimation}

\label{subsec:error_estimation}

\begin{figure}[t]
     \centering
     \includegraphics[width=\linewidth]{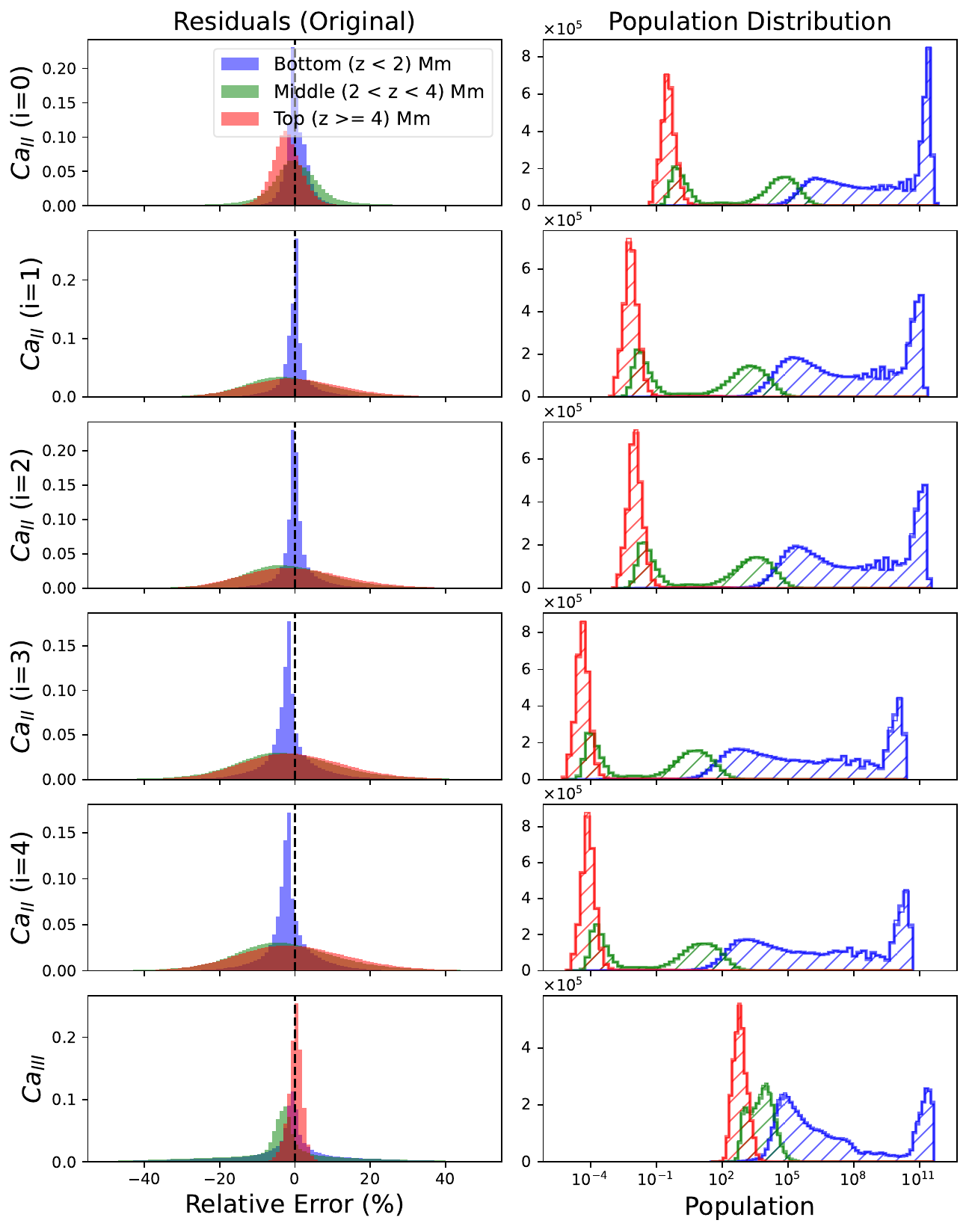}
     \caption{Left panels show the error distributions for predicted \ion{Ca}{ii} populations, stratified by atmospheric regime.
     blue: lower heights ($z<2$ Mm); green: mid-chromosphere ($2 \leq z< 4$ Mm); red: upper layers ($z\geq4$ Mm).
     Right panels show the distribution of the density of those populations. Solid lines outline the population of the 3D NLTE solution, and striped areas show the inferred distribution from the GNN solution.}
     \label{fig:histogram_errors}
\end{figure}

To rigorously assess the source of these discrepancies, we stratified the atmosphere into three distinct regimes (lower, middle, and top) based on the height $z$.
Each regime corresponds to a region where different ionization stages dominate (e.g., \ion{Ca}{ii} vs \ion{Ca}{iii}).

The histograms in Fig.~\ref{fig:histogram_errors} quantify the error distribution across these heights.
We observe the following trends:
\begin{itemize}
    \item Lower regime ($z<2$~Mm): The network achieves very good precision.
    The physics here is dominated by LTE or weak non-LTE effects, which are easier to capture.
    \item Middle regime ($2$~Mm $ \leq z < 4$~Mm): This region, corresponding roughly to the chromosphere where \ion{Ca}{ii} lines are formed, shows a broader error distribution but remains well-centered around zero.
    \item Top regime ($z\geq4$~Mm): The errors are larger here, consistent with the transition to the corona where \ion{Ca}{ii} populations are negligible. Accordingly, this is where \ion{Ca}{iii} errors are the lowest.
\end{itemize}

\begin{figure*}[t]
    \centering 
    \includegraphics[width=0.75\textwidth]{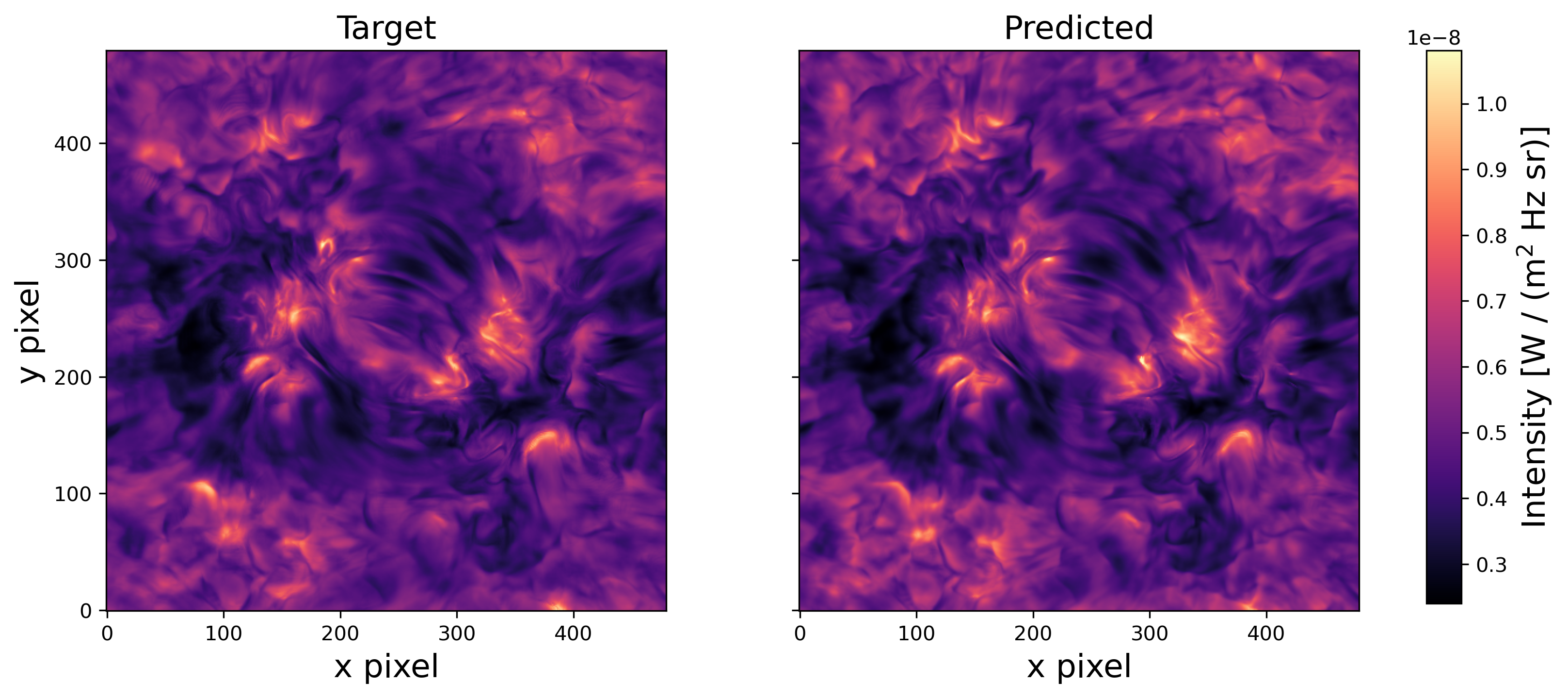}
    \includegraphics[width=0.75\textwidth]{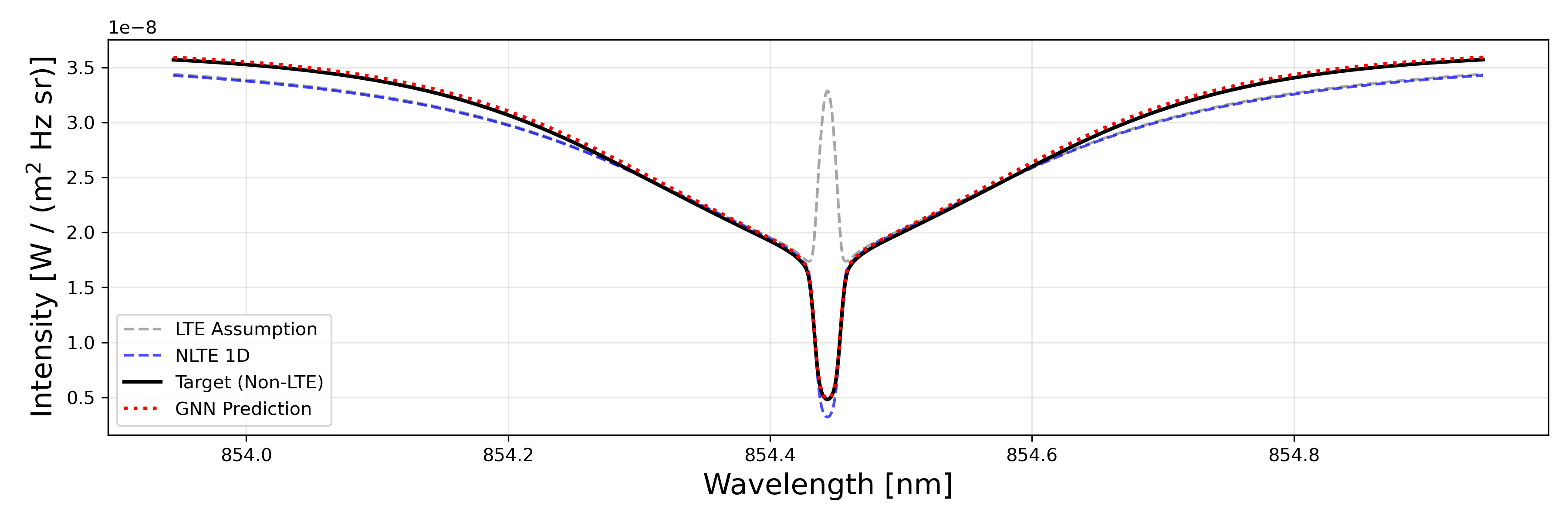}
    \caption{Synthesized maps at the center of the \ion{Ca}{ii} 8542 \AA\ line for the \texttt{Multi3D}, predicted, LTE and 1D non-LTE populations.
    The second last row shows the mean profile over the synthesized region for the \texttt{Multi3D}, the predicted, and the LTE/non-LTE populations.
    }
    \label{fig:synthesis_maps}
\end{figure*}

Crucially, the histograms demonstrate that in the normalized space (which the network optimizes directly), the error distributions remain Gaussian across all regimes, confirming that the network has converged to a robust solution.

\subsection{Comparison between different strides}
\label{subsec:Muiltistride}

To evaluate how node separation influences prediction accuracy, we utilized the dynamic connection stride capability of our graph construction (as detailed in Sect. \ref{sec:training}).
We perform inference on the same test map using fixed stride values within the window $S \in \{1,2,3,4,5\}$ and compared the resulting performance.
Note that we use strides larger than those used during training, and the network generalizes well.
Although some outliers are found with relative errors up to $\sim50\%$, we checked and those relative errors may be misleading since they correspond to the bottom left part of the scatter plots in Fig. \ref{fig:inference_scatter}, where the populations are low, and therefore small deviations (in absolute terms) can lead to a very high relative error estimations.

The analysis yielded the following observations:
\begin{itemize}
    \item {Contextual improvement}: Increasing the stride and, consequently, the physical context, improved the accuracy of the prediction up to $S\sim4$.
    \item {Diminishing returns}: The most significant performance improvement was observed when increasing the stride from $S=1$ to $S=2$, while subsequent increases provided diminishing returns.
    \item {Performance limit}: At $S=5$, the results worsened slightly but with minimal differences. We expect to find worse results if
    $S$ is increased too much because the sampling around the column of interest becomes too sparse.
\end{itemize}

We conclude that optimal performance is achieved with $N=4$ selected neighbors at a stride of $S=4$.
This configuration allows the network to effectively capture a context of $\pm 16$ columns, corresponding to a physical area of approximately $\sim1.5\times 1.5$~Mm$^2$, roughly 2 arcsec$^2$.
We acknowledge that this behavior can be biased by the training strategy, where only strides from $S=1 \to 5$ were introduced. Obviously, including different number of neighbors and different strides can potentially lead to different outcomes.
However, we note that non-local effects are expected to have a limited action distance (limited by when one finds $\Delta \tau > 1$ at horizontal directions), so that including further pixels from the central column might become irrelevant.
A precise determination of this distance would require computing optical depths throughout the full simulation
cube in all directions, a computationally demanding task that we consider an important avenue
for other work.
As a reference, the photon mean free path in the lower solar chromosphere is of order $\sim 100$ km \citep{Judge_2015}.
In the horizontal direction, photons could potentially travel over larger distances before being absorbed, as the chromosphere density is lower and the variation is less in the horizontal direction.
Nevertheless, the performance of our model saturates empirically for context radii $R\sim1500$ km ($>10$ vertical mean free paths), suggesting that the dominant contribution of horizontal non-local effects on the populations is captured within this scale, even if not all horizontal radiative coupling is fully enclosed.

\section{Line synthesis}
\label{sec:synthesis}

\begin{figure*}[t]
    \centering
    \includegraphics[width=0.75\textwidth]{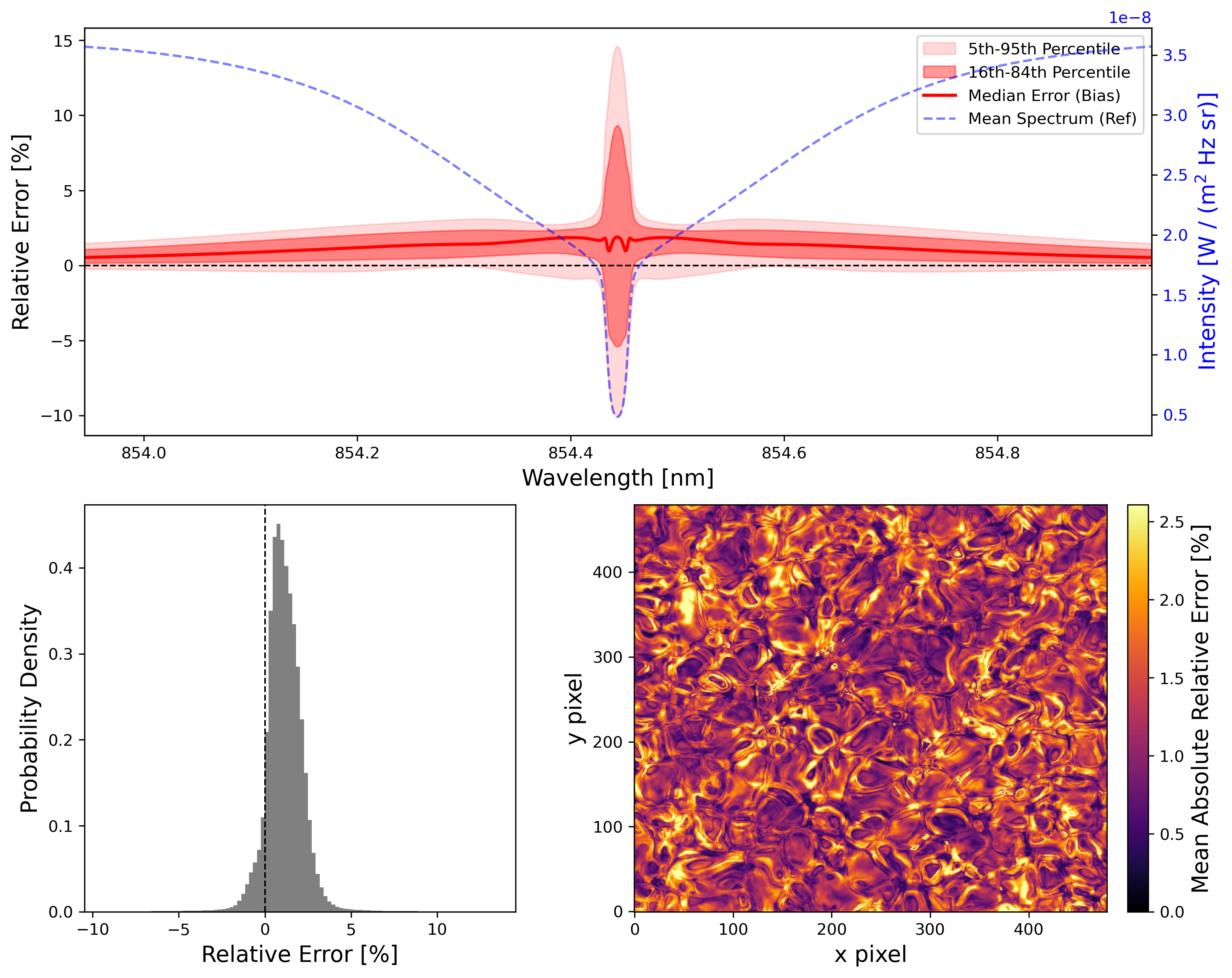}
    \caption{Error distribution of the synthetic \ion{Ca}{II} 8542 \AA\ line.
    First panel is the wavelength distribution, with percentiles 5, 16, 84, 95 with a double axis on the right side representing the mean line profile over the test set.
    Second row shows the histogram of the error distribution (wavelength aggregated) of the synthetic profiles and a map of the mean error across all the wavelengths on every pixel of the synthesized map.}
    \label{fig:synthesis_errors}
\end{figure*}

The ultimate validation of this method lies in demonstrating that predicted populations yield spectra indistinguishable from those computed with full 3D radiative transfer. Because our inferred populations already encode the essential 3D non-LTE physics—such as horizontal scattering and geometric effects—computing the emergent spectrum requires only a single formal solution of the radiative transfer equation along the line of sight, entirely bypassing the iterative SE equations. To demonstrate this, we synthesized the \ion{Ca}{ii}~8542\,\AA\ spectral line for our test set using the \texttt{Lightweaver} framework \citep{lightweaver}.

The resulting monochromatic maps at the core of the line (upper left panel for the target value and upper right panel for the 
predicted) are shown in Fig.~\ref{fig:synthesis_maps} alongside the mean profiles (lower panel).
The monochromatic maps look almost identical both in appearance and in their intensity values across all wavelengths. Some small
differences can be found, though, in the regions with more intensity. The mean profile obtained with the GNN almost perfectly
overlaps with the target mean profile. Both are different from the equivalent non-LTE profile computed in a 1.5D approximation (i.e., column by column).
For comparison, we also show the LTE profile, which shows the typical emission core produced by the increased temperature
in chromospheric layers.

To quantify our errors further, Fig.~\ref{fig:synthesis_errors} presents the error distribution of the synthesized spectra both across wavelengths (upper
panel) and spatially (lower right panel).
The errors introduced by the GNN approximation are negligible, with $>99\%$ of the intensities having relative errors below $5\%$ (see lower left panel of
Fig.~\ref{fig:synthesis_errors}). It is clear that the errors are larger in the line core, where more variability is expected. However, most of the points 
are still recovered with $<10\%$ relative error with 68\% probability.
Concerning the spatial distribution of the errors, although some spatial clustering is visible, the overall magnitude remains small and shows 
no clear correlation with specific morphological features, suggesting the errors are not systematically biased by atmospheric structure.
This analysis confirms that the 3D GNN is a viable surrogate for expensive 3D RT calculations in the context of forward modeling and inversions.

\section{Conclusions}
\label{sec:conclusions}
We have presented a novel approach to significantly accelerate 3D non-LTE radiative transfer problems using Graph Neural Networks.
By discretizing the solar atmosphere as a 3D graph (where nodes encode local physical conditions and edges represent radiative coupling) we trained a model that efficiently predicts the statistical equilibrium populations of the \ion{Ca}{ii} atom from local thermodynamic parameters and their spatial context.

The GNN architecture successfully captures the 3D spatial dependencies inherent in the non-LTE problem. Our ``star-like'' connectivity scheme allows efficient aggregation of information from surrounding columns while avoiding the prohibitive memory cost of a fully connected graph.
The dynamic striding technique employed to construct the graph also allowed us to make the technique robust for different grid sizes, potentially expanding its usability in observational data.
The model achieves high fidelity in reproducing populations computed by \texttt{Multi3D} on the \bifrost\ simulation, with excellent performance in the photosphere and mid-chromosphere.
A slight degradation in the upper chromosphere (where non-local effects are strongest) is expected, but the errors remain unbiased and Gaussian-distributed.

Critically, the inference speed is estimated to be $\sim10^6$ faster than traditional iterative solvers\footnote{This is a conservative estimation assuming $250$ kh for the 3D NLTE solution and $\sim 10$ min for the GNN inference. Inference can be as fast as 2 min, and 3D NLTE solutions can take up to millions of CPU h.}.
This acceleration is essential for future inversion codes that aim to incorporate 3D radiative transfer effects.
Furthermore, spectral synthesis of the \ion{Ca}{ii}~8542\,\AA\ line using predicted populations yields intensity profiles in excellent agreement with the ground truth, with residual errors well below typical observational noise.

An important caveat of this method is that the simulation, while state-of-the-art, is constrained by specific initial and boundary conditions that may not capture phenomena such as flaring regions, highly dynamic jets, or quiet-Sun conditions at their extremes.
Deploying the model for real inversions would therefore require training on a more diverse set of simulations sampling a broader parameter space.
We note that, in our experience from previous work with graph neural networks \citep{Andreu_2022}, these architectures tend to be remarkably robust and can generalize well beyond the strict boundaries of their training data.
Nevertheless, we would not recommend their use for conditions significantly outside the training distribution without first validating performance on representative examples.
We view the extension of training or testing with more diverse training sets as an important avenue, since it requires a complete 3D NLTE solution to accomplish.

This work opens the possibility of integrating this approach of forward modeling into an inversion framework.
Our approach enables a strategy analogous to \texttt{DeSIRe} or \texttt{Hazel} \citep{asensio_trujillo_hazel08} for 3D: by predicting the 3D atomic populations (or departure coefficients in the case of \texttt{DeSIRe}) from a model atmosphere, the non-LTE inversion problem can be effectively reduced to an LTE-like optimization.
The GNN acts as a fast surrogate for the SE equations, allowing standard inversion codes to retrieve 3D atmospheric properties by simply modifying the source function and opacity, thereby bypassing the prohibitive cost of full 3D radiative transfer.
This architecture can also be extended to predict full Stokes synthesis.
Finally, expanding the training set to include energetic events from 3D flare simulations \citep[e.g.,][]{Cheung2019,Hansteen2015} will be crucial for generalization to active region conditions.

Another avenue of improvement is to generalize it to the polarized case. In the atomic density matrix formalism, the diagonal 
elements $\rho^0_0$ represent the level populations, while off-diagonal elements $\rho^K_{Q\neq 0}$ encode atomic 
polarization induced by anisotropic radiation or magnetic fields. Our current architecture predicts only the diagonal elements 
of the density matrix, $\rho^0_0$ (i.e., the level populations), which enables the synthesis of Stokes~$I$. Extending our
model, arguably including magnetic fields, to predict the off-diagonal elements $\rho^K_{Q\neq K}$ could open the
possibility of performing full-Stokes polarized non-LTE radiative transfer including the Zeeman splitting and 
quantum interferences (Hanle effect). We leave this idea for future work.

Finally, we also leave for the future the synthesis of lines with strong partial redistribution effects. This would require recomputing the populations using this formalism and retraining the model. Once the populations are computed, the formal solution will be more time-consuming because it needs to take into account the redistribution function. However, huge gains can still be obtained with respect to the full calculation. 

\begin{acknowledgements}
      AVA acknowledges support from the Deutsche Forschungsgemeinschaft (DFG), project number 538773352.
      AAR acknowledges funding from the Agencia Estatal de Investigación del Ministerio de Ciencia, Innovación y Universidades (MCIU/AEI) under grant 
      ``Polarimetric Inference of Magnetic Fields'' and the European Regional Development Fund (ERDF) with reference PID2022-136563NB-I00/10.13039/501100011033. 
      Part of this research was also supported by the Research Council of Norway through its Centres of Excellence scheme, project number 262622.
      We thank the kind advice and contributions from Chris Osborne with the usage of the \texttt{Lightweaver} framework.
      We also kindly appreciate the insights on the synthesis process provided by Tanaus\'u del Pino Alem\'an.
      We acknowledge the community effort devoted to the development of the following open-source packages: \texttt{PyTorch} \citep{PyTorch}, \texttt{PyTorch Geometric}, \texttt{NumPy}, and \texttt{Matplotlib}. This research has made use of NASA's Astrophysics Data System Bibliographic Services.      
\end{acknowledgements}

\bibliographystyle{aa} 
\bibliography{aa.bib} 

\begin{appendix}

\clearpage

\end{appendix}
\end{document}